\documentclass{PoS}

\newcommand{\be}{\begin{equation}}
\newcommand{\ee}{\end{equation}}
\newcommand{\bea}{\begin{eqnarray}}
\newcommand{\eea}{\end{eqnarray}}

\newcommand{\nn}{\nonumber}

\newcommand{\bra}{\langle}
\newcommand{\ket}{\rangle}
\newcommand{\eps}{\epsilon}
\newcommand{\vareps}{\varepsilon}
\newcommand{\re}{\mbox{Re\,}}
\newcommand{\im}{\mbox{Im\,}}

\newcommand{\rmR}{{\rm R}}
\newcommand{\rmI}{{\rm I}}

\newcommand{\pv}{{\mathbf p}}

\title{Can complex Langevin dynamics evade the sign problem?}

\ShortTitle{Can complex Langevin dynamics evade the sign problem?}

\author{\speaker{Gert Aarts}%
        \\
	Physics Department, Swansea University, Swansea, United Kingdom\\
        E-mail: \email{g.aarts@swan.ac.uk}}

\abstract{
 I answer the question in the title for the 
relativistic Bose gas at finite chemical potential using 
numerical lattice simulations, complemented with analytical understanding.
 }

\FullConference{The XXVII International Symposium on Lattice Field Theory\\
		 July 26-31, 2009\\
		 Peking University, Beijing, China}

\begin{document}

\section{Introduction}

As is well known, at nonzero baryon chemical potential the complexity of 
the fermion determinant prohibits the use of importance sampling in 
lattice QCD. This makes the determination of the QCD phase diagram an 
outstanding open problem (a summary of what is possible despite this 
obstruction can be found in Ref.\ \cite{deF}). Stochastic quantization 
\cite{Parisi:1980ys} does not rely on importance sampling: the 
configurations that dominate in the partition function are found by 
integrating complex Langevin equations in an enlarged phase space, after a 
complexification of the degrees of freedom 
\cite{Parisi:1984cs,Klauder:1983,Damgaard:1987rr}. During last year's 
Conference I reported \cite{Aarts:2008tc} on work done in collaboration 
with Ion-Olimpiu Stamatescu on QCD with static quarks and related one-link 
models \cite{Aarts:2008rr}. In the one-link models we found excellent 
results for all values of the chemical potential, from zero all the way to 
saturation.  Moreover, the sign problem was present but clearly not 
dangerous.  The results in the heavy dense limit of QCD were preliminary 
but encouraging.

These results triggered a number of intriguing questions. For instance,
can complex Langevin dynamics handle
 \begin{itemize}
\item  a severe sign problem?
\item  the thermodynamic limit?
\item  phase transitions?
\item  the Silver Blaze problem  \cite{Cohen:2003kd}?
\end{itemize}

I will answer these questions positively in a theory that is much simpler 
than QCD: the relativistic Bose gas at nonzero chemical potential 
\cite{Aarts:2008wh}. Even though it lacks the intricacies of a nonabelian 
gauge theory, it is a four-dimensional lattice field theory with a sign 
problem, a Silver Blaze problem, and a phase transition. As an additional 
benefit, one can study the complex Langevin equations perturbatively, 
allowing detailed analytical insight into the algorithm 
\cite{Aarts:2009hn}. Of course, there are many more questions to ask and 
work to address those is currently in progress.

\section{Relativistic Bose gas}

The lattice action is 
\be
 S = \sum_x \bigg[ \left(2d+m^2\right) \phi_x^*\phi_x
 + \lambda\left( \phi_x^*\phi_x\right)^2
- \sum_{\nu=1}^4\left(  \phi_x^* e^{-\mu\delta_{\nu,4}} \phi_{x+\hat\nu}
+ \phi_{x+\hat\nu}^* e^{\mu\delta_{\nu,4}} \phi_x \right)
\bigg],
\ee
 where $d=4$ and we take $m^2>0$. It satisfies $S^*(\mu)=S(-\mu^*)$, just 
as in QCD. In continuum notation, the nonderivative part of the Lagrangian 
density reads $\left(m^2-\mu^2\right)|\phi|^2+\lambda|\phi|^4$: at tree 
level, a phase transition at $\mu_c=m$ separates the vacuum (symmetric) 
and the high-density (symmetry-broken) phases. The $\mu$-independence of 
bulk thermodynamic quantities as long as $\mu<\mu_c$, even though the 
action, the Boltzmann weight, and hence the field configurations depend 
explicitly on $\mu$, is the Silver Blaze problem \cite{Cohen:2003kd}. 
Equating the imaginary part of the action with zero yields the phase 
quenched theory: this is a theory with a $\mu$-dependent mass parameter 
and no sign problem.

After decomposing $\phi=\left(\phi_1+i\phi_2\right)/\sqrt{2}$ and 
complexifying $\phi_a\to \phi_a^{\rm R}+ \phi_a^{\rm I}$ ($a=1,2$), it 
is straightforward to write down the complex Langevin equations for the 
(now) four real fields,
 \be
 \frac{\partial \phi_a^{\rm R}}{\partial\theta} =
        -\re\,\frac{\delta S}{\delta\phi_a}\Big|_{\phi_a\to\phi_a^{\rm R}
+ i\phi_a^{\rm I}} +\eta_a, 
\;\;\;\;\;\;\;\;\;\;\;\;\;\;\;\;
  \frac{\partial \phi_a^{\rm I}}{\partial\theta} =
  -\im\, \frac{\delta S}{\delta\phi_a}\Big|_{\phi_a\to\phi_a^{\rm R} +
i\phi^{\rm I}},
\ee
 where $\theta$ is the Langevin time and $\eta_a$ is Gaussian noise with 
$\bra\eta_a\eta_b\ket=2\delta_{ab}$.
 It is also straightforward to solve these equations numerically 
\cite{Aarts:2008wh}. I have used $m=\lambda=1$ on lattices of size 
$\Omega=N^4$, with $N=4,6,8,10$, and Langevin time step $\eps=5\times 
10^{-5}$ and not encountered instabilities in the numerical integration. 
The phase quenched theory is solved using real Langevin dynamics (with 
$\phi^\rmI\equiv 0$).

\begin{figure}[h]
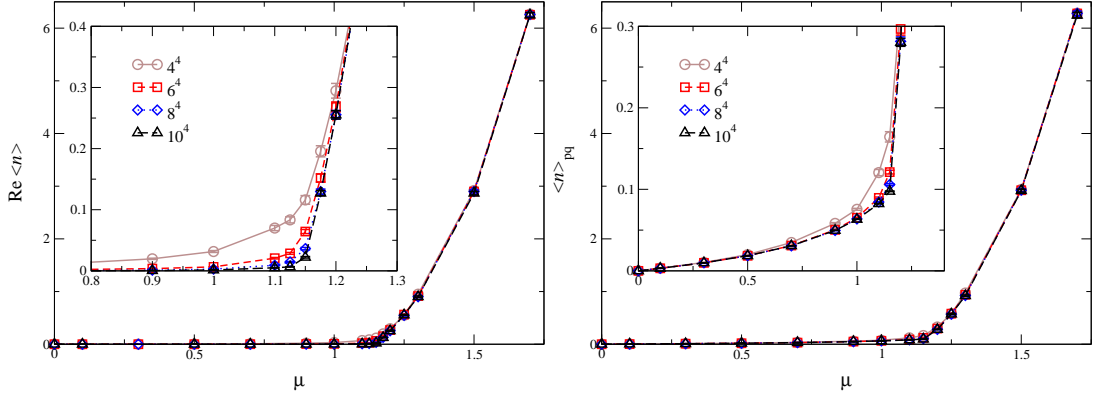

\vspace*{0.3cm}
\begin{center}
\includegraphics[height=5.2cm]{fig12.eps}
\includegraphics[height=5.2cm]{fig4.eps}
\end{center}
 \caption{
 Density in the full (left) and the phase quenched (right) theory as a 
function of chemical potential on lattices with size $N^4$, with 
$N=4,6,8,10$ ($m=\lambda=1$).
 }
\label{fig:1}
\end{figure}

The resulting density $\bra n\ket=\Omega^{-1}\partial\ln Z/\partial\mu$ is 
shown in Fig. \ref{fig:1}, in the full theory (left) and the phase 
quenched theory (right).  The Silver Blaze feature in the full theory is 
immediately visible. In the thermodynamic limit the density is strictly 
zero as long as $\mu<\mu_c\approx 1.15$.  The phase quenched theory has no 
Silver Blaze region: instead the density increases as soon as $\mu$ is 
nonzero. The $\mu$-independence in the full theory is a direct result of 
the complexity of the action and the sign problem.

In this model the complex Langevin equations can be studied perturbatively 
\cite{Aarts:2009hn}. Ignoring interactions, the solution of these 
equations reads, in momentum space and in the Silver Blaze region only,
 \bea
\phi_{a,p}^{\rm R}(\theta) &=& e^{-A_p\theta}\left[\cos (B_p\theta) 
\phi_{a,p}^{\rm R}(0) +i\sin(B_p\theta) \vareps_{ab}\phi_{b,p}^{\rm 
I}(0)\right]
\nn \\ && 
+\int_0^\theta ds\, 
e^{-A_p(\theta-s)}\cos[B_p(\theta-s)]\eta_{a,p}(s),
\\
\phi_{a,p}^{\rm I}(\theta) &=& e^{-A_p\theta}\left[\cos (B_p\theta) 
\phi_{a,p}^{\rm I}(0) -i\sin(B_p\theta) \vareps_{ab}\phi_{b,p}^{\rm 
R}(0)\right]
\nn \\
&& -i\int_0^\theta ds\, 
e^{-A_p(\theta-s)}\sin[B_p(\theta-s)]\vareps_{ab}\eta_{b,p}(s).
\eea
 Here $\phi_{a,p}^{\rm R,I}(0)$ are the initial conditions, 
$\vareps_{ab}$ 
is the totally antisymmetric tensor with $\eps_{11}=1$, and the 
coefficients $A_p$, $B_p$ are
 \bea
A_p &=& m^2 + 4\sum_{i=1}^3\sin^2\frac{p_i}{2}  + 2 \left( 1-
\cosh \mu \cos p_4\right)
\to m^2-\mu^2+ \pv^2 + p_4^2, \\
B_p &=& 2 \sinh\mu \sin p_4 \to  2\mu p_4.
\eea
 The expressions after the arrows correspond to the formal continuum 
limit. This solution is only valid when $A_p>0$ (or $\mu<\mu_c$), which 
reflects the standard instability of the free Bose gas. When $\mu>\mu_c$, 
symmetry breaks and the inclusion of the $\lambda|\phi|^4$ term is 
necessary for stabilization.

\begin{figure}[h]
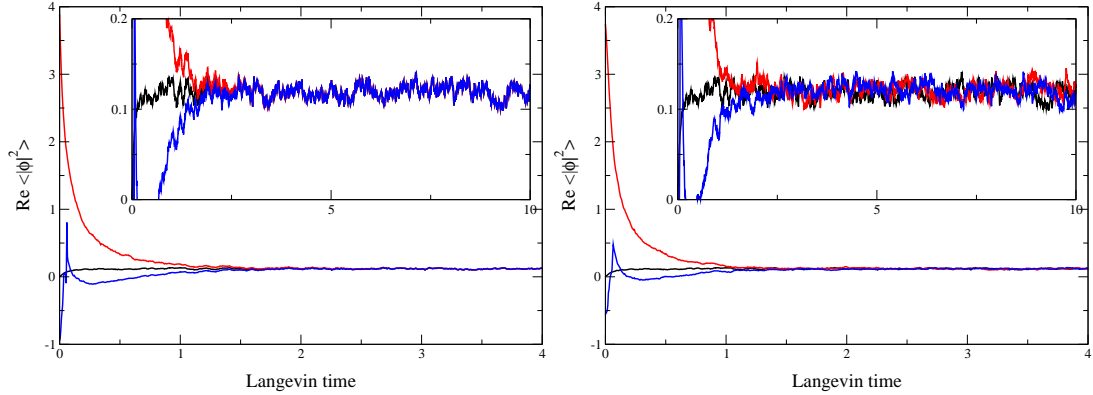

\vspace*{0.3cm}
\begin{center}
\includegraphics[height=5.2cm]{plot_hist_rephi_full_4x4x4x4_mu0.5_ss.eps}
\includegraphics[height=5.2cm]{plot_hist_rephi_full_4x4x4x4_mu0.5_ds.eps}
\end{center}
 \caption{
 Independence of initial conditions: Langevin evolution of 
$\bra|\phi|^2\ket$ with three different 
initial conditions using the same (left) and a different (right) random 
number sequence ($\mu=0.5$, $m=\lambda=1$, $N^4=4^4$).
 }
\label{fig:ic}
\end{figure}

From the analytical solution, we find that both the independence of 
initial conditions and convergence at large Langevin time are controlled 
by $e^{-A_p\theta}$. This is demonstrated numerically for the interacting 
theory in Fig.\ \ref{fig:ic} for $\mu=0.5$. The left figure shows the 
Langevin history of $\bra|\phi|^2\ket$, using three different initial 
conditions but the same random number sequence in the stochastic process. 
Since $\mu=0.5$ is far from the critical value (and hence the minimal 
value of $A_p$ is well separated from zero), the memory of initial 
conditions is quickly erased. 
To have an indication for the size of the statistical fluctuations during 
the evolution, runs with a different random number sequence are shown in 
the figure on the right. Note that in a typical run the Langevin time 
$0<\theta<250$.

 Expectation values are obtained by noise averaging and taking the limit 
$\theta\to \infty$. Provided again that $A_p>0$, we find from the 
analytical solution
 \bea
\lim_{\theta\to\infty}
\bra\phi^\rmR_{a,-p}(\theta)\phi^\rmR_{b,p'}(\theta)\ket &=&
 \delta_{ab} \delta_{pp'} 
\frac{1}{2A_p}\frac{2A_p^2+B_p^2}{A_p^2+B_p^2},
\nn\\
\lim_{\theta\to\infty}
\bra\phi^\rmI_{a,-p}(\theta)\phi^\rmI_{b,p'}(\theta)\ket &=&
  \delta_{ab} \delta_{pp'} \frac{1}{2A_p}\frac{B_p^2}{A_p^2+B_p^2},
\nn\\
\lim_{\theta\to\infty}  
\bra\phi^\rmR_{a,-p}(\theta)\phi^\rmI_{b,p'}(\theta)\ket &=&
  \vareps_{ab} \delta_{pp'} \frac{i}{2}\frac{B_p}{A_p^2+B_p^2}.
\eea
 So far interactions have been neglected in this analytical calculation. 
An elegant way to include interactions on the mean field level is by 
looking for fixed points of the Langevin equations for the two-point 
functions above, using a Gaussian factorization 
\cite{Aarts:2009hn}. After the dust settles, the result is a simple shift 
in the effective mass parameter, $m^2 \to m^2+4\lambda\bra|\phi|^2\ket$, 
as expected. The critical value of the chemical potential in the 
mean field approximation then follows from 
$m^2+4\lambda\bra|\phi|^2\ket+2(1-\cosh\mu)=0$, where $\bra|\phi|^2\ket$ 
is determined by a self-consistent gap equation. For the parameters used 
here we find $\mu_c=1.15$, in agreement with the numerical estimate.

\begin{figure}[t]
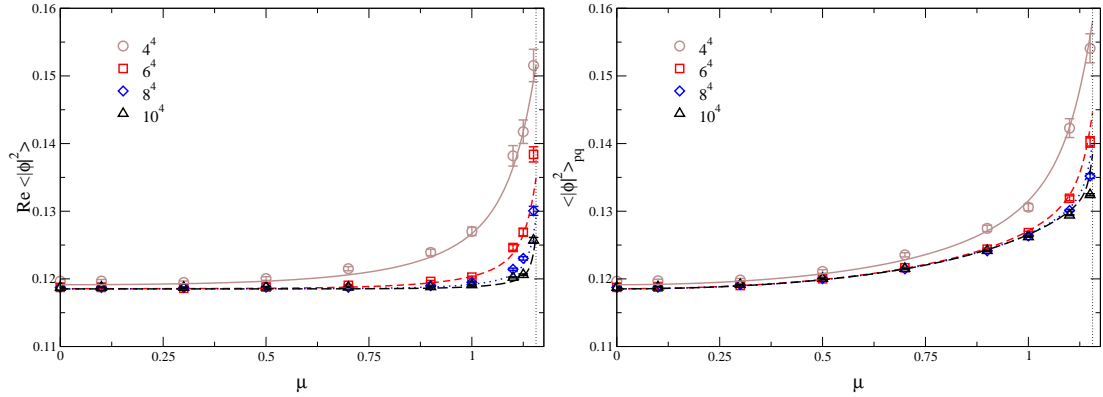

\vspace*{0.3cm}
\begin{center}
\includegraphics[height=5.2cm]{ana_full_phi2_MF.eps}
\includegraphics[height=5.2cm]{ana_phq_phi2_MF.eps}
\end{center}
 \caption{
 Comparison between the mean field predictions (lines) and the numerical 
data (symbols)  for $\bra|\phi|^2\ket$ in the full (left) and phase 
quenched (right) theories in the Silver Blaze region. The vertical dotted 
line indicates the mean field estimate for the critical chemical 
potential.
 }
\label{fig:ana1}
\end{figure}

\begin{figure}[b]
\begin{center}
\includegraphics[height=5.2cm]{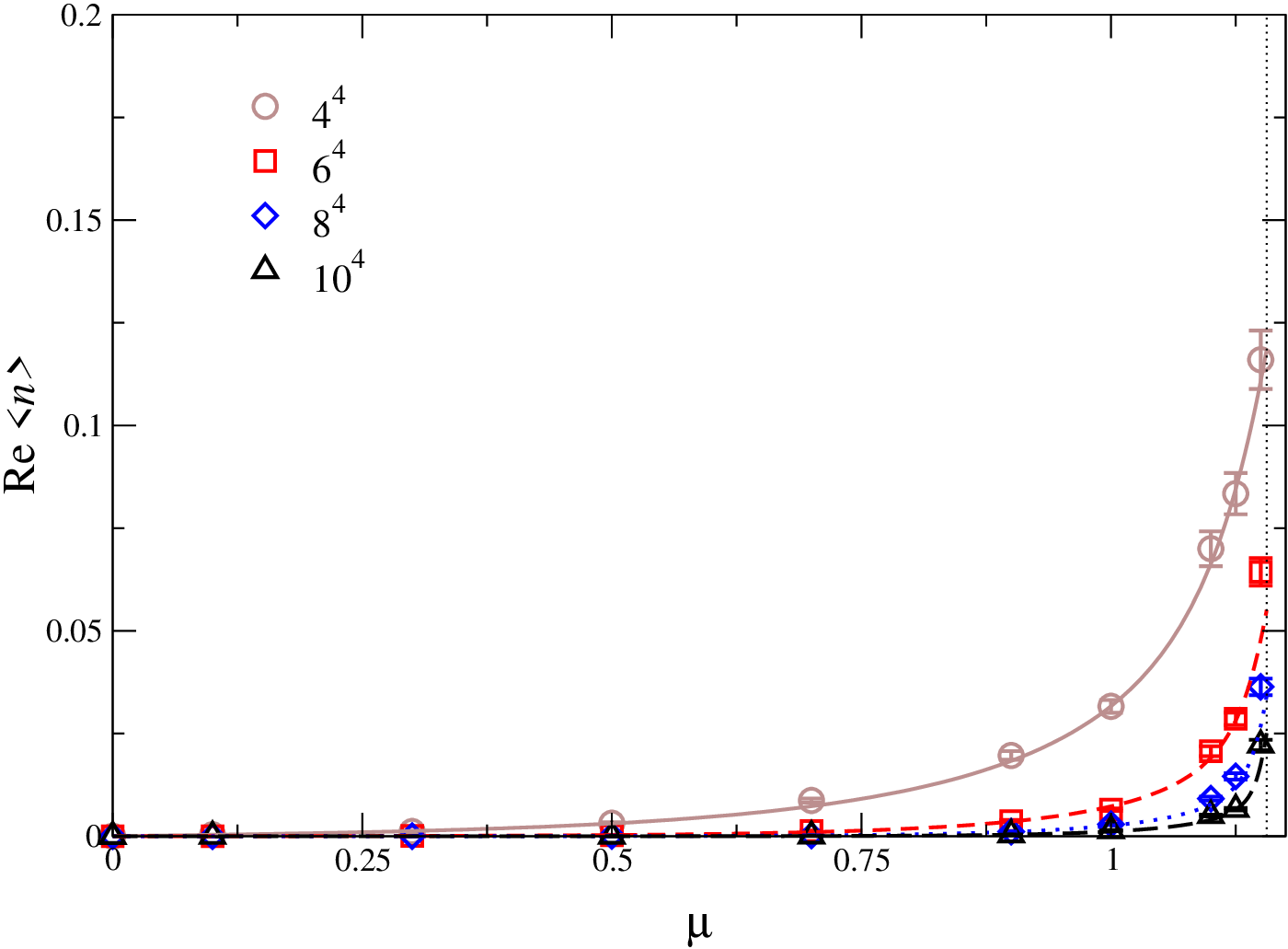}
\includegraphics[height=5.2cm]{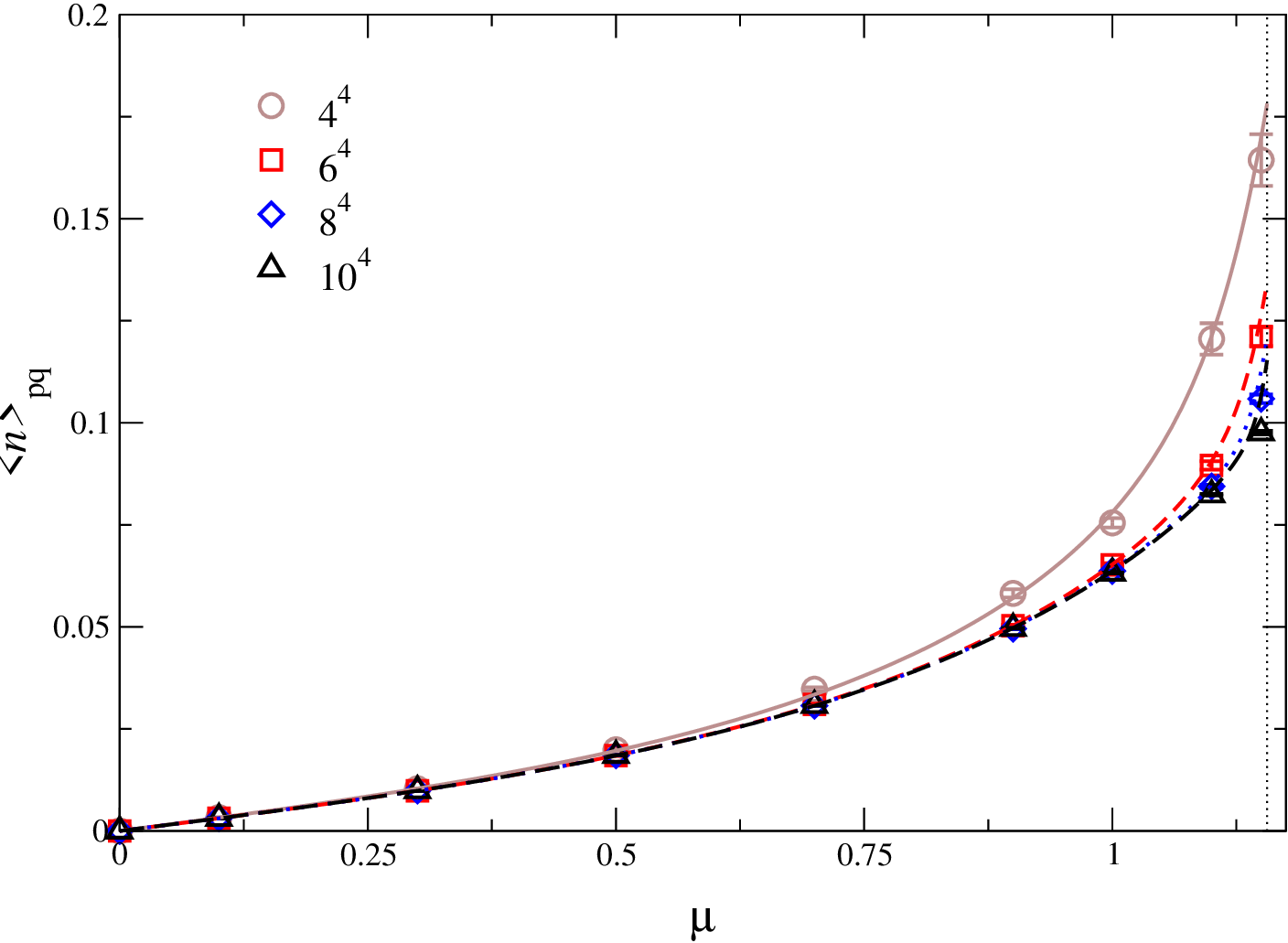}
\end{center}
 \caption{As above, for the density  $\bra n\ket$.}
\label{fig:ana2}
\end{figure}

We can now compare the nonperturbative results from the numerical solution 
of the complex Langevin equations with the mean field estimates on a 
finite lattice. This is shown in Fig.\ \ref{fig:ana1} for 
$\bra|\phi|^2\ket$ and in Fig.\ \ref{fig:ana2} for $\bra n\ket$. Since at 
$\mu=0$ the action is real, the value of $\bra|\phi|^2\ket$ at $\mu=0$ is 
obtained with real Langevin dynamics. This provides a nontrivial check of 
the numerical code. We observe excellent agreement between the simulations 
and the analytical results, including finite-size effects. This indicates 
that complex Langevin dynamics is successful, but also that the mean field 
approximation is adequate and the theory is effectively weakly coupled.

\begin{figure}[t]
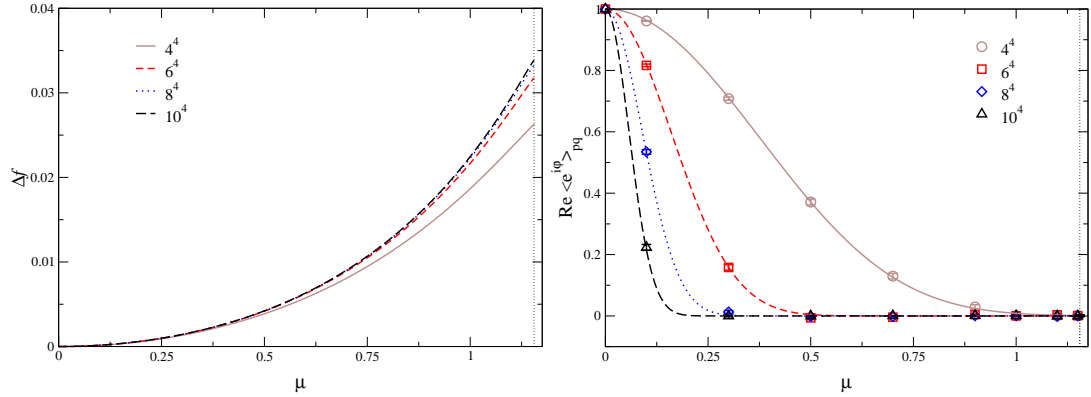

\vspace*{0.3cm}
\begin{center}
\includegraphics[height=5.2cm]{ana_Deltaf_MF.eps}
\includegraphics[height=5.2cm]{ana_avphase_MF.eps}
\end{center}
 \caption{
 Left: difference $\Delta f$ between the free energy densities of the full 
and the phase quenched theories in the mean field approximation. Right: as 
in the preceding figures, for the average phase factor $\bra 
e^{i\varphi}\ket_{\rm pq}$ in the phase quenched theory.
 }
\label{fig:ana3}
\end{figure}

The results so far answer three of the questions raised in the 
Introduction. What about the severity of the sign problem? From the 
pronounced difference between the full and the phase quenched theory, one 
can expect the sign problem to be severe. This is quantified by studying 
the average phase factor,
\be
 \bra e^{i\varphi}\ket_{\rm pq},
\;\;\;\;\;\;\;\;\;\;\;\;\;\;\;\;
 e^{i\varphi} = \frac{e^{-S}}{|e^{-S}|}, 
\ee
 where the expectation value is taken in the phase quenched theory, with 
real weight $|e^{-S}|$. This phase factor is relevant for reweighting and 
is expected to vanish exponentially in the thermodynamic limit, since
\be 
 \bra e^{i\varphi}\ket_{\rm pq} = \frac{Z_{\rm full}}{Z_{\rm pq}} = 
e^{-\Omega\Delta f},
\ee
 where $\Delta f$ is the difference between the free energy densities in 
the full and the phase quenched theories. The mean field results for the 
difference between the free energies are shown in Fig.\ \ref{fig:ana3} 
(left). We note a rapid convergence in the thermodynamic limit. Inserting 
these results for $\Delta f$ in the exponential $e^{-\Omega\Delta f}$ 
yields the average phase factor. In the numerical simulations the phase 
factor is measured directly. A comparison between the two is shown in 
Fig.\ \ref{fig:ana3} (right). We observe again excellent agreement. The 
severeness of the sign problem is manifest. As mentioned above, this is 
necessary and does not affect complex Langevin dynamics in practice.

\section{Outlook}

The question posed in the title and made more specific in the Introduction 
has been answered positively for the relativistic Bose gas at finite 
chemical potential considered here. In this model the sign problem is 
required to yield the correct physics, summarized by the Silver Blaze 
problem, and it is severe. Nevertheless, complex Langevin dynamics can be 
applied without encountering (numerical) problems affecting e.g.\ 
stability or convergence. The outcome of the numerical simulations can be 
understood from a comparison with mean field theory. Moreover, the 
inner workings of the algorithm can be analysed as well, relying 
essentially on the weakly coupled nature of this theory.

Does this mean that stochastic quantization can easily be applied to other 
theories with a complex action due to a nonvanishing chemical potential? 
We are currently studying complex Langevin dynamics in a variety of 
models, including the three-dimensional XY model at nonzero chemical 
potential, and extending our analysis of QCD with static quarks. At this 
stage it is fair to say that {\em easily} is slightly too optimistic.

\acknowledgments

Discussions with Philippe de Forcrand, Simon Hands, Frank James, Erhard 
Seiler, Kim Splittorff and Ion-Olimpiu Stamatescu are greatly appreciated.
I am grateful to the Royal Society for Conference support.

\end{document}